\documentclass[conference, letterpaper]{IEEEtran} %

\usepackage{cite}
\usepackage{tabularx}
\usepackage{subcaption} 
\usepackage{amsmath,amssymb,amsfonts}
\usepackage{graphicx}
\usepackage{comment}
\usepackage{float}
\usepackage{hyperref}
\usepackage{multirow}
\usepackage{booktabs} 
\usepackage{siunitx}
\usepackage{tabularray}
\usepackage{textcomp}
\usepackage{tikz}
\usepackage{amsmath}
\usepackage{amsfonts}
\usepackage{adjustbox}
 \usepackage{url} 
\usepackage{multirow}
\usepackage{xcolor}
\usepackage{pgfplots}
\usepackage{multirow} 

\usepackage{subcaption}
\usepackage[affil-it]{authblk} 
\usepackage{booktabs}
\usepackage{algpseudocode}
\usepackage[ruled, lined, longend, linesnumbered]{algorithm2e}
\usepackage{amsmath}
\usepackage[utf8]{inputenc}
\usepackage{tikz}

\usepackage{amssymb}
\usepackage{pifont}

\begin{document}

\title{Zero-Day Botnet Attack Detection in IoV: A Modular Approach Using Isolation Forests and Particle Swarm Optimization}

\author[2]{Abdelaziz Amara korba}
\author[1]{Nour Elislem Karabadji}
\author[2]{Yacine Ghamri-Doudane}

\affil[1]{National Higher School of Technology and Engineering,  LTSE, E3360100, Annaba, Algeria.}

\affil[2]{L3I, University of La Rochelle, France}

\maketitle

\begin{abstract}
The Internet of Vehicles (IoV) is transforming transportation by enhancing connectivity and enabling autonomous driving. However, this increased interconnectivity introduces new security vulnerabilities. Bot malware and cyberattacks pose significant risks to Connected and Autonomous Vehicles (CAVs), as demonstrated by real-world incidents involving remote vehicle system compromise. To address these challenges, we propose an edge-based Intrusion Detection System (IDS) that monitors network traffic to and from CAVs. Our detection model is based on a meta-ensemble classifier capable of recognizing known (N-day) attacks and detecting previously unseen (zero-day) attacks. The approach involves training multiple Isolation Forest (IF) models on Multi-access Edge Computing (MEC) servers, with each IF specialized in identifying a specific type of botnet attack. These IFs, either trained locally or shared by other MEC nodes, are then aggregated using a Particle Swarm Optimization (PSO) based stacking strategy to construct a robust meta-classifier. The proposed IDS has been evaluated on a vehicular botnet dataset, achieving an average detection rate of 92.80\% for N-day attacks and 77.32\% for zero-day attacks. These results highlight the effectiveness of our solution in detecting both known and emerging threats, providing a scalable and adaptive defense mechanism for CAVs within the IoV ecosystem.
\end{abstract}


\section{Introduction}

The Internet of Vehicles (IoV) refers to a network of interconnections between vehicles and transportation infrastructure using advanced communication technologies. The IoV has significantly advanced transportation by enhancing connectivity, safety, and convenience, enabling real-time traffic updates, improved navigation, and autonomous driving. However, this increased connectivity also introduces new security vulnerabilities \cite{boualouache20235g}, making CAV and autonomous vehicles susceptible to cyber threats such as cyberattacks and malwares, which can compromise safety and data integrity. Among the seven possible threats related to a vehicle’s data and code identified by the UNECE WP 29.R155   \cite{turgeman_2022} is the introduction of malware. Over 900 cyberattacks on vehicles have been reported publicly in the past decade . One example includes an incident where attackers remotely took control of a vehicle’s engine while it was operating on a highway.

Although securing V2X networks \cite{boualouache2023reinforcement} and detecting intrusions in IoV systems \cite{korba2023federated,zero-x,VDoS} have attracted considerable attention in recent years, the detection of malware attacks in vehicular environments remains relatively underexplored \cite{haghighi2023cyber,antibot}, despite its critical importance and potential impact. In this work, we focus on bot malware, which infects a device and transforms it into a "bot" under the control of a malicious actor. These compromised devices form a botnet, which can be used for attacks such as DDoS, data theft, and spam distribution. Bot malware poses a serious threat to CAVs and autonomous vehicles by disrupting critical systems and compromising passenger safety. This threat is already well-known in the field of the Internet of Things (IoT), where numerous devices have been compromised to conduct large-scale attacks, such as the infamous Mirai botnet \cite{Nokia2023ThreatReport}.

Artificial Intelligence (AI) techniques, particularly Machine Learning (ML) and Deep Learning (DL), have significantly enhanced the effectiveness of existing IDSs~\cite{labiod2022fog,diaf2024beyond,diaf2025bartpredict} by expanding the range of detectable threats and improving both efficiency and precision. However, current ML-based IDSs face several limitations in effectively securing the IoV environment. First, in a real IoV environment, zero-day attacks are frequent, yet most ML-based IDSs \cite{antibot, boualouache2022federated} are designed for static and closed-set scenarios. They assume all possible attack types are known and predefined during training, which is unrealistic due to the constant emergence of new attacks. Supervised learning, commonly used in current IDSs, is inefficient at identifying unknown attacks with novel patterns. Second, training ML models requires large datasets, which are hard to obtain due to privacy concerns. Real-world scenarios also present class imbalances, where attacks occur with varying frequencies across the IoV ecosystem. This imbalance leads to models performing well on frequently seen attack types but poorly on less common ones. Third, deploying IDSs at the CAV level poses challenges in data collection, computational capacity for ML model training, and integration into proprietary vehicle systems. Intellectual property restrictions further complicate deployment within onboard systems, making it difficult to implement effective ML-based IDS solutions directly within CAVs.

Our solution addresses the challenge of zero-day attacks by proposing a new Meta-Ensemble classifier. We train multiple Isolation Forests (IFs)~\cite{liu2008isolation}, each specialized in recognizing a specific type of attack, and then combine them into an ensemble using a PSO-based stacking strategy. If all the IFs identify an instance as an anomaly, it is considered an unknown (zero-day) attack. This modular approach mitigates the problems of data availability and class imbalance. If one IF performs poorly, it does not adversely affect the overall performance of the meta-classifier in recognizing other types of attacks. Additionally, it is possible to combine locally trained IFs with those trained elsewhere, expanding the range of detectable attacks. Our IDS monitors all network traffic exchanged between the CAV and the Internet. By executing at the edge, specifically on Multi-access Edge Computing (MEC) servers, our IDS enables efficient data collection and leverages the computational resources of MEC servers to train and run ML models.

The remainder of this paper is organized as follows: Section~\ref{RT} reviews related work. Section~\ref{TM} presents the threat model. The proposed methodology is detailed in Section~\ref{SOL}. Section~\ref{SIM} illustrates the performance evaluation results, and finally, Section~\ref{CON} concludes the paper.

\section{Related Work} \label{RT}

Ensuring the security of the Internet of Vehicles (IoV) has become a critical challenge due to its highly dynamic and interconnected nature. Intrusion Detection Systems (IDSs) have emerged as a fundamental defense mechanism, with most solutions relying on AI-driven behavioral analysis to detect cyber threats \cite{zero-x, korba2024ai, korba2024life, korba2023federated}. Various advanced AI paradigms have been explored to enhance the adaptability and accuracy of IDSs, including federated learning \cite{korba2023federated} to ensure privacy, semi-supervised learning and open-set classification to address unknown attacks \cite{zero-x, korba2023federated}, continual learning to enable adaptive threat detection \cite{korba2024life}.

Botnet detection in traditional and IoT networks, especially after the emergence of the Mirai IoT botnet, has gained significant interest in recent years. Wei et al. \cite{wei2023lightweight} developed a deep learning framework for early-stage IoT botnet detection, notable for its use of a 5-second detection window and packet payload-independent features, enhancing timely anomaly identification. Nguyen et al. \cite{nguyen2022collaborative} investigated collaborative machine learning models for early IoT botnet detection, evaluating algorithms like Support Vector Machine and K-Nearest Neighbors. Bojarajulu et al. \cite{bojarajulu2023intelligent} introduced SMIE (Slime Mould with Immunity Evolution), optimizing a hybrid classifier with Bidirectional Gated Recurrent Units (Bi-GRU) and Recurrent Neural Networks (RNN), marking a significant advancement in botnet detection accuracy.

A recent study \cite{haghighi2023cyber} introduced a covert attack on vehicle electronic modules via malware in intelligent transportation systems. This attack subtly manipulates module operations to increase accident rates without alarming human observers. The study focused on the Adaptive Cruise Controller (ACC) module and proposed a non-disruptive, add-on Intrusion Detection System (IDS) to defend against such attacks. The IDS described in the study focuses on monitoring intra-vehicle communications within the CAN bus, rather than network traffic.

Garip et al. \cite{garip2019shieldnet} developed SHIELDNET, a machine learning-based detection system targeting the GHOST botnet communication protocol, which uses Basic Safety Messages (BSMs). SHIELDNET detects anomalies in BSM fields unique to vehicular networks. However, its effectiveness is constrained by its dependency on the GHOST protocol, making it less adaptable to changes in the botnet's communication methods. AntibotV \cite{antibot}, a multilevel behavior-based framework monitors vehicle activity at both network and in-vehicle levels, using decision tree classifiers trained on collected network traffic data. Notably, this work also studied bot IoV attack scenarios and developed a realistic new dataset. However, AntibotV relies on closed-set and supervised learning, limiting its ability to detect zero-day attacks. 

In \cite{korba2023federated}, a federated deep autoencoder-based anomaly detection approach has been proposed for detecting attacks in V2X networks. This approach allows for the detection of unknown attacks but does not facilitate the recognition of specific types of detected attacks (N-day). In \cite{zero-x}, a blockchain-enabled federated open set learning approach is introduced to detect zero-day attacks. Both approaches consider V2V and V2I attack scenarios but do not address attacks by bot malware. Additionally, the IDS operates at the vehicle level, which may present processing challenges related to the training of ML models.

Karabadji et al.~\cite{karabadji2023accuracy} proposed a multi-objective genetic algorithm to construct optimized Random Forests by jointly considering accuracy, diversity, and model compactness. Their method selects a subset of accurate and diverse trees using a binary chromosome representation and a fitness function promoting generalization. Applied to IoT security, it enhances botnet detection by improving ensemble robustness and interpretability. In a subsequent study, they introduced a graph-guided variant that constrains the search space to relevant features, achieving strong results in IoV scenarios~\cite{karabadji2024genetic}. However, both approaches rely on supervised learning, which limits their ability to detect zero-day attacks due to the lack of labeled data.

Despite these efforts, research on detecting vehicular bot malware in the IoV through network traffic analysis remains limited. Further research is needed to develop mechanisms capable of detecting zero-day bot attacks. These mechanisms should focus on V2N communication channels, which are used for communication with remote hackers, and ensure that the proposed scheme takes into account the resource limitations of CAVs, particularly regarding data collection and the training of ML models.

\begin{figure*}[ht!]
    \centering
    \includegraphics[width=0.68\textwidth]{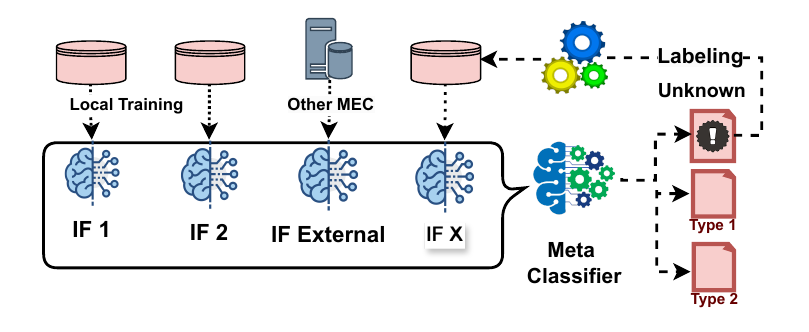}
    \caption{Meta-Ensemble Network Traffic Classifier}
    \label{fig:workflow}
\end{figure*}

\section{Threat Model}  \label{TM}
We consider attack vectors executed by bot malware targeting to compromise CAVs. When a CAV is compromised by a bot malware,it becomes part of a botnet controlled by a botmaster (hacker) remotely through a C2 (Command and Control) server. While we believe that all attacks applicable by botnets in IoT networks are possible, we think the following three categories of cyber threats are the most relevant in the context of IoV \cite{antibot}:
\begin{itemize}
    \item \textbf{Eavesdropping}: The hackers attempt to intercept private communications without authorization.  An example of such an attack was demonstrated by \cite{checkoway2011comprehensive}, who found that audio surveillance inside vehicles is possible. In their research, they showed that attackers could exploit virtual assistance tools, such as Siri, by embedding malicious commands in recorded music, innocuous-sounding speech, or even low-powered lasers. These commands can trick Siri into recording conversations using the vehicle’s internal microphones and sending the captured audio to a remote server controlled by the hacker every.

\item \textbf{GPS Tracking}: The hacker may attempt to track the vehicle's location in real-time, check the location, or retrieve the complete trajectory. For real-time tracking, the GPS information must be sent every second (streaming) to the botmaster. The information can also be sent periodically or on-demand. In this paper, we consider the real-time GPS tracking scenario. The bot vehicle sends the latitude, longitude, speed, time, and direction coordination in streaming to the botmaster.

\item \textbf{Denial of Service (DoS)}: Vehicles can either be victims of DoS attacks or be used to launch such attacks. Rahal et al. \cite{antibot} investigated DoS attacks exploiting the WAVE Short Message Protocol (WSMP), which is critical for vehicle safety communications. These attacks, including WSMP flood and geographic WSMP flood, overwhelm vehicle resources by sending forged packets with unknown PSIDs, preventing legitimate message processing. These attacks can prevent the transfer of safety messages between vehicles, potentially causing catastrophic damage.
\end{itemize}


\section{Proposed Approach} \label{SOL}
We propose an Edge-enabled Intrusion Detection System (IDS) to monitor network traffic to and from Connected Autonomous Vehicles (CAVs) passing through the base station. Our approach utilizes a modular detection method that involves training multiple Isolation Forests (IFs) at each Multi-access Edge Computing (MEC) server, with each IF trained to recognize a specific type of botnet attack. These IFs are then combined using a Particle Swarm Optimization (PSO) based stacking strategy to form a Meta-ensemble classifier. In the following sections, we detail our proposed approach.

\subsection{Edge-Based Network Traffic Monitoring}
To address the challenges related to implementing the IDS at the CAV level, we deploy the IDS on the Multi-access Edge Computing (MEC) server located at the base station where all traffic transits. This ensures that the perspective of the network traffic exchanged with the Internet is the same as it would be when monitoring at the CAV level. Our IDS is composed of a network traffic collector, which collects network traffic data, a data processing module that filters and preprocesses the collected traffic, and a ML-based detection model that analyzes the network traffic to identify bot attacks. A crucial aspect of our system is the continuous update of the IF models. Whenever a new attack is detected by a MEC, a new IF model is trained for that specific attack and shared through a secure channel with other MECs. This allows all MEC servers to quickly benefit from new insights and maintain their detection capabilities up to date against evolving threats. The establishment of the secure channel falls outside the scope of this paper.

\subsection{Meta-Ensemble Network Traffic Classifier}
Our method trains multiple IFs, each targeting a specific type of attack. For each attack \( c_i \) in a dataset \( D \), a dedicated Isolation Forest \( IF_i \) is trained to recognize that attack and detect anomalies in other behaviors. Figure \ref{fig:workflow} illustrates the functioning of the Meta-Ensemble classifier. This training process leverages network traffic data to capture the unique characteristics and distribution of each attack effectively. Each Isolation Forest, composed of multiple trees denoted as $Tree_{1}, Tree_{2}, ..., Tree_{n}$, works together to identify anomalies within their respective class. Within each Isolation Forest, the trees vote on whether a data point is normal (+1) or an anomaly (-1). 

The pseudocode for the Isolation Forest algorithm~\cite{liu2008isolation} is provided in Algorithm \ref{algo:psiforest} (Lines 1 - 7). 

For a given sample \( x \), the anomaly score \( s_i(x) \) by \( IF_i \) is defined as follows:

%

\begin{equation}
s_i(x) = 2^{-\frac{E(h(x))}{c(n)}}
\end{equation}
where \( E(h(x)) \) is the mean path length of \( x \) and \( c(n) \) is a normalization constant given by:
\begin{equation}
c(n) = 2H(n-1) - \frac{2(n-1)}{n}
\end{equation}
with \( H(i) \) being the \( i \)-th harmonic number, approximated by \( \ln(i) + \gamma \) (where \( \gamma \) is the Euler-Mascheroni constant).

\begin{algorithm}
\caption{PSO-Optimized Meta-iForest Classifier}
\label{algo:psiforest}
\KwIn{$X$: input data, $T$ : number of trees, $S$: sub-sampling size, $P$: particle population size, $G$: number of generations}
\KwOut{$\mathcal{M}$: A meta-ensemble classifier \textit{iForest}}

\textbf{Step 1: Isolation Forest Construction} \\
Initialize an empty forest $\mathcal{F}$\;
Set height limit $L = \lceil \log_2 S \rceil$\;
\For{$i \gets 1$ \KwTo $T$}{
    Draw a random sample $X' \sim \text{sample}(X, S)$\;
    $\mathcal{F} \gets \mathcal{F} \cup \text{iTree}(X', 0, L)$\;
}

\textbf{Step 2: PSO-Based Thresholds Optimization} \\
Initialize particle population with random thresholds\;
\For{$g \gets 1$ \KwTo $G$}{
    \ForEach{particle $p$ in the swarm}{
        Evaluate fitness $F(p)$ using ensemble accuracy and detection rate\;
        Update personal best $P_{best}$ and global best $G_{best}$ thresholds\;
         Update particle velocity and position using Eq~\ref{eq:pso1}
    }
}

\textbf{Step 3: Meta-Ensemble Classifier Construction} \\
Combine iForest models $\mathcal{F}$ with thresholds ${\theta_i}$ to form the meta-classifier
$\mathcal{M}$:
\[
\mathcal{M} = \sum_{i=1}^{T} \alpha_i \cdot \text{iForest}(X_i', \theta_i)
\]
where $\alpha_i$ denotes the performance-based weight of each iForest\;
\Return $\mathcal{M}$\;

\end{algorithm}

These IFs are then combined using a Particle Swarm Optimization (PSO)~\cite{hadjadji2024optimizing} based stacking strategy to form a meta-classifier. PSO optimizes anomaly thresholds for IFs by considering their collective performance, unlike independent optimization which can lead to suboptimal results. The goal of PSO is to enhance the overall detection accuracy and sensitivity of the ensemble by leveraging synergistic effects, avoiding local optima, and managing complex interactions. We encode the set of anomaly thresholds \( \mathbf{X}_p = [t_1, t_2, \ldots, t_n] \) as particle. Each decision variable \( t_i \) corresponds to the threshold for class \( c_i \). Each particle in the swarm represents a candidate solution in the form of a vector of thresholds \( \mathbf{X}_p = [t_1, t_2, \ldots, t_n] \). The particle updates follow these equations:
\begin{equation}
\label{eq:pso1}
\begin{split}
V^{(i)}_{d}(t+1) = & \, wV^{(i)}_{d}(t) + c_1 r_1 \left(P^{(i)}_{d} - X^{(i)}_{d}(t)\right) \\
& + c_2 r_2 \left(G^{(i)}_{d} - X^{(i)}_{d}(t)\right)
\end{split}
\end{equation}

\begin{equation}
X^{(i)}_{d}(t+1) = X^{(i)}_{d}(t) + V^{(i)}_{d}(t+1)
\end{equation}

where \( V^{(i)}_{d} \) is the velocity of the \( i \)-th particle in dimension \( d \), \( X^{(i)}_{d} \) is the position, \( P^{(i)}_{d} \) is the personal best solution, \( G^{(i)}_{d} \) is the global best solution, \( w \) is the inertia weight, and \( c_1 \) and \( c_2 \) are acceleration coefficients with \( r_1 \) and \( r_2 \) being uniformly distributed random numbers.

In order to find the best candidate solution, we define a fitness function $\mathcal{F}(p)$ that evaluates two key criteria: the accuracy of the ensemble of Isolation Forests (IFs) for a given set of anomaly thresholds, and the ability to detect unknown attacks. We use the normal (benign) traffic to adjust the anomaly thresholds to detect unknown attacks. Here, the normal traffic serves as an unknown traffic class to evaluate the robustness of the meta-ensemble classifier. The fitness function is defined as follows: \begin{equation} \mathcal{F}(p) = \dfrac{\text{Accuracy} + \text{Detection}}{2} \end{equation}

where the accuracy is calculated as follow:
\begin{equation}
\text{Accuracy} = \frac{\text{Number of Correctly Classified Instances}}{\text{Total Number of Instances}}
\end{equation}
and the detection of the normal class by the ensemble is calculated as the rate of instances known to be normal that will not be recognized by the ensemble of Isolation Forests, using the following equation:
\begin{equation}
\text{Detection} = \frac{\text{Number Unknown Instances}}{\text{Total Number of  Normal Instances}}
\end{equation}


The values of the decision variables \( t_i \) from each particle are used as thresholds for the corresponding Isolation Forests \( IF_i \). These thresholds determine the sensitivity of each Isolation Forest to detecting anomalies in the corresponding class. An instance is classified into class \( I_i \) if its anomaly score \( s_i(x) \) is less than or equal to the anomaly threshold \( \theta_i \) of class \( i \), i.e., \( s_i(x) \leq \theta_i \). In the case where an instance is classified into two or more classes, the class with the smallest anomaly score is assigned, i.e., the class \( I_j \) such that \( s_j(x) = \min\{s_k(x) \mid s_k(x) \leq \theta_k \} \). If all the IFs recognize an instance as an anomaly, then it is considered an unknown (zero-day) attack. After defining the thresholds \( t_i \), the accuracy of the ensemble of Isolation Forests is evaluated by classifying each instance \( i \) in the test set. As shown in Algorithm \ref{algo:psiforest}, the construction of the meta-classifier involves three key steps: building multiple iForest models, optimizing their thresholds using PSO, and combining the optimized models to form a highly effective ensemble classifier.

\section{Performance Evaluation}\label{SIM}
In this section, we outline the methodology used to assess the effectiveness of our approach. We first present the evaluation conducted on the AntibotV dataset~\cite{antibot}, which includes various botnet attacks in the IoV. 

\subsection{Dataset Processing \& Experimental Setup}
To test the proposed approach, we utilized the dataset developed in our previous research \cite{antibot}, which includes malicious network traffic from four botnet attacks: Eavesdropping, WSMP Flood, Geographic WSMP Flood, and GPS Tracking. For further information on these attacks, please refer to our article \cite{antibot}. The dataset comprises a collection of PCAP files. 

\begin{table}[!htbp]
\caption{Packet and Flow Statistics}
\centering
\renewcommand{\arraystretch}{1.2} 
\begin{tabularx}{\columnwidth}{@{}lXrr@{}} 
    \toprule
    \textbf{Traffic} & \textbf{Type} & \textbf{Packets} & \textbf{Flows} \\ 
    \midrule
    \multirow{2}{*}{\textbf{Benign}} 
        & Normal IP & 282107 & 1150 \\ 
        & WSMP Traffic & 18544 & 364 \\ 
    \midrule
    \multirow{4}{*}{\textbf{Malicious}} 
        & GPS Tracking & 22502 & 429 \\ 
        & Eavesdropping & 29372 & 320 \\ 
        & WSMP Flood & 1385017 & 143 \\ 
        & Geo WSMP Flood & 16131 & 24 \\ 
    \bottomrule
\end{tabularx}
\label{tab:packet_flow_stats}
\end{table}

We test our approach by representing the network traffic both as a set of packets and as a set of flows. Therefore, we created two datasets: one packet-based and one flow-based. For flow extraction and feature calculation, we developed scripts using two distinct traffic exporters: CICFlowMeter~\cite{CICFlowMeter} for extracting network flows and calculating features, and Tranalyzer~\cite{tranalyzer} for packet-based feature calculation. The features under consideration can be categorized into four main types: time-based, byte-based, packet-based, and protocol-based (network and transport layers). For more details on the flow-based features, the reader is referred to~\cite{CICFlowMeter}, while information on the packet-based features can be found in~\cite{tranalyzer}. The distribution of samples by attack type, following data preprocessing operations such as cleaning and normalization, is presented in Table~\ref{tab:packet_flow_stats}.

Our solution was implemented in Python 3 on the Google Colab cloud environment. We used the Scikit-learn~\cite{scikit} package to implement the Isolation Forest (IF) models. To create a realistic test scenario, we designated one type of attack as the 0-day attack while considering the remaining attack types as N-day attacks. This process was repeated with different 0-day attacks, resulting in K test scenarios, where K represents the number of attack types in the dataset. For each test scenario, the system was trained with N-1 attack types, ensuring that one attack type was not seen during training (0-day attack). We conducted a 5-fold cross-validation to ensure the robustness of our models. The evaluation of our proposed approach was based on the following performance metrics: 
\begin{itemize}
    \item Precision: \quad $\displaystyle {TP}/{(TP + FP)}$
    \item Recall: \quad $\displaystyle {TP}/{(TP + FN)}$
    \item F1-Score: \quad $\displaystyle 2 \times {(\text{Precision} \times \text{Recall})} / {(\text{Precision} + \text{Recall})}$
    \item Detection Rate (0-day):  \quad $\displaystyle {TP}/{(TP + FN)}$
\end{itemize}

TP, TN, FP, and FN denote true positive, true negative, false positive, and false negative, respectively.

\begin{table*}[t]
    \vspace*{3mm} 
    \caption{Packet-based Predictive Performances}
    \centering
    \renewcommand{\arraystretch}{1.2}
    \begin{tabularx}{\textwidth}{>{\raggedright\arraybackslash}p{4cm}|
    X@{\hspace{0.3cm}}X@{\hspace{0.3cm}}
    X@{\hspace{0.3cm}}X@{\hspace{0.3cm}}
    X@{\hspace{0.3cm}}X|
    X@{\hspace{0.3cm}}X}
        \toprule
        & \multicolumn{6}{c|}{\textbf{N-Day Scenarios}} & \multicolumn{2}{c}{\textbf{0-Day Scenarios}} \\
        \cmidrule(lr){2-7} \cmidrule(lr){8-9}
        \textbf{Metrics} & \multicolumn{2}{c}{Precision (\%)} & \multicolumn{2}{c}{Recall (\%)} & \multicolumn{2}{c|}{F1-Score (\%)} & \multicolumn{2}{c}{Detection Rate (\%)} \\
        \cmidrule(r){1-1} \cmidrule(lr){2-3} \cmidrule(lr){4-5} \cmidrule(lr){6-7} \cmidrule(l){8-9}
        \textbf{Approach} & Naive & PSO & Naive & PSO & Naive & PSO & Naive & PSO \\
        \cmidrule(r){1-1} \cmidrule(lr){2-2} \cmidrule(lr){3-3}
        \cmidrule(lr){4-4} \cmidrule(lr){5-5}
        \cmidrule(lr){6-6} \cmidrule(lr){7-7}
        \cmidrule(l){8-8} \cmidrule(l){9-9}
        Eavesdropping     & 58.58 & 66.94 & 77.33 & 71.67 & \textbf{66.66} & \textbf{69.22} & 17.90 & 40.94 \\
        GPS Tracking      & 65.90 & 47.12 & 53.83 & 65.00 & \textbf{59.26} & 54.63 & 4.52   & 55.20 \\
        WSMP Flood        & \textbf{100.00} & \textbf{99.66} & 86.00 & \textbf{93.83} & \textbf{92.47} & \textbf{96.66} & \textbf{99.60} & \textbf{99.54} \\
        Geo WSMP Flood    & 89.74 & 97.19 & \textbf{90.50} & 80.50 & \textbf{90.12} & \textbf{88.06} & 6.50   & 65.23 \\
        \bottomrule
    \end{tabularx}
    \label{tab:perPacket}
\end{table*}

\begin{table*}[t]
    \vspace*{3mm}
        \caption{Overall Performances Comparison (Flow-based)}

    \centering
    \renewcommand{\arraystretch}{1.2}
    \begin{tabularx}{\textwidth}{>{\raggedright\arraybackslash}p{4cm}|
    X@{\hspace{0.3cm}}X@{\hspace{0.3cm}}
    X@{\hspace{0.3cm}}X@{\hspace{0.3cm}}
    X@{\hspace{0.3cm}}X|
    X@{\hspace{0.3cm}}X}
        \toprule
        & \multicolumn{6}{c|}{\textbf{N-Day Scenarios}} & \multicolumn{2}{c}{\textbf{0-Day Scenarios}} \\
        \cmidrule(lr){2-7} \cmidrule(lr){8-9}
        \textbf{Metrics} & \multicolumn{2}{c}{Precision (\%)} & \multicolumn{2}{c}{Recall (\%)} & \multicolumn{2}{c|}{F1-Score (\%)} & \multicolumn{2}{c}{Detection Rate (\%)} \\
        \cmidrule(r){1-1} \cmidrule(lr){2-3} \cmidrule(lr){4-5} \cmidrule(lr){6-7} \cmidrule(l){8-9}
        \textbf{Approach} & Naive & PSO & Naive & PSO & Naive & PSO & Naive & PSO \\
        \cmidrule(r){1-1} \cmidrule(lr){2-2} \cmidrule(lr){3-3}
        \cmidrule(lr){4-4} \cmidrule(lr){5-5}
        \cmidrule(lr){6-6} \cmidrule(lr){7-7}
        \cmidrule(l){8-8} \cmidrule(l){9-9}
        Eavesdropping   & \textbf{95.56} & 90.63 & 59.90 & \textbf{100.00} & 73.64 & \textbf{95.08} & \textbf{42.50} & 30.83 \\
        GPS Tracking    & 89.32 & \textbf{98.86} & 73.68 & \textbf{93.73} & 80.75 & \textbf{96.23} & 30.00 & \textbf{92.44} \\
        WSMP Flood      & 97.78 & \textbf{98.03} & 67.23 & \textbf{90.34} & 79.68 & \textbf{94.03} & 40.00 & \textbf{100.00} \\
        Geo WSMP Flood  & 42.32 & \textbf{93.33} & \textbf{93.33} & 86.67 & 58.23 & \textbf{89.88} & 16.50 & \textbf{86.01} \\
        \bottomrule
    \end{tabularx}
    \label{tab:perFlow}
\end{table*}

\begin{figure*}
    \centering
    \includegraphics[width=1\textwidth]{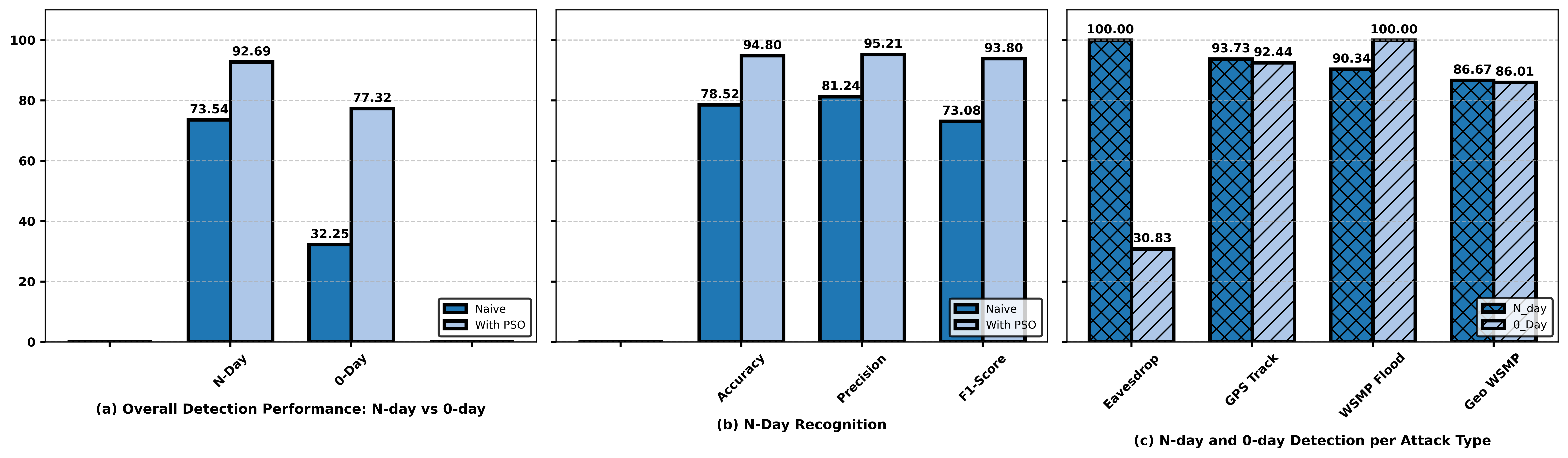}
    \caption{Overall Performances Comparison (Flow-based)}
    \label{fig:compa}
\end{figure*}

\subsection{Experimental results}
The experimental results obtained by analyzing network traffic at both the packet and flow levels are presented in Tables \ref{tab:perPacket} and \ref{tab:perFlow}, respectively. Each row in these tables provides the detection performance for each attack, considering both 0-day and N-day scenarios. Additionally, we compare our approach with a naive baseline that employs a fixed anomaly threshold for all Isolation Forests (IFs), allowing us to highlight the benefits of our optimization strategy.

Table \ref{tab:perPacket} illustrates the results of the packet-based detection approach. With the naive configuration, the system performs well for WSMP Flood and Geo WSMP Flood in the N-day scenario, but struggles with Eavesdropping and GPS Tracking. For the 0-day scenario, WSMP Flood shows high detection rates, while Eavesdropping and GPS Tracking are significantly weak. Introducing PSO enhances performance, particularly for zero-day attacks, significantly boosting detection rates for Eavesdropping and GPS Tracking. However, while it improves F1-Scores for WSMP Flood and Geo WSMP Flood in N-day scenarios, it does not adequately address the lower performance for Eavesdropping and GPS Tracking. Overall, PSO improves zero-day attack detection but requires further optimization to consistently enhance performance across all attack types.

Table~\ref{tab:perFlow} presents the results of the flow-based detection approach. In N-day scenarios, the proposed method achieves substantial improvements across key metrics. The integration of PSO consistently enhances Recall and F1-Score, reflecting more reliable recognition of known attacks. For instance, Recall for the Eavesdropping attack increased from 59.9\% to 100\%, resulting in an F1-Score improvement from 73.64\% to 95.08\%. Similar gains are observed for GPS Tracking, WSMP Flood, and Geo WSMP Flood, where the application of PSO improves F1-Scores, confirming the effectiveness of the proposed approach in identifying all four attack types.

When evaluating zero-day attacks, the PSO-based approach significantly improves the detection rate across several attack types. For instance, the detection rate for \textit{GPS Tracking} increased substantially from 30\% to 92.44\%, and for \textit{WSMP Flood}, from 40\% to 100\%. However, our solution exhibits a low detection rate for the \textit{Eavesdropping} attack in the zero-day scenario, both with the naive approach and with PSO. This limitation can be attributed to the stealthy nature of Eavesdropping traffic, which makes it difficult to distinguish from benign communication. Overall, these results underscore the robustness of our approach in not only recognizing N-day attacks but also in effectively detecting zero-day attacks.

Figure~\ref{fig:compa} provides a comparative overview of the performance of our proposed approach. Figure~2.a compares the average (macro) detection rate of our method with that of the naive baseline. The results show a clear improvement, particularly in the detection of zero-day attacks. Figure~2.b presents a comparison of accuracy, average precision, and average F1-score, again demonstrating that our approach consistently outperforms the naive method across all metrics. Figure~2.c summarizes detection performance per attack type, under both N-day and 0-day scenarios. The results indicate strong performance overall, with the exception of the eavesdropping attack in the 0-day context. This limitation can be attributed to the nature of eavesdropping, which is a passive attack where the adversary listens to network communications without actively interfering. As a result, the corresponding traffic patterns closely resemble those of legitimate traffic, making detection particularly difficult. The stealthy behavior of such attacks allows them to evade anomaly-based mechanisms that rely on observable deviations in traffic features. This highlights the inherent challenge of detecting passive threats like eavesdropping, which blend seamlessly into normal network activity.

The results obtained demonstrate that our approach performs well in flow-based detection mode. Conversely, packet-based analysis is effective primarily for detecting DoS attacks and struggles with more sophisticated threats such as GPS tracking and eavesdropping. Our flow-based solution exhibits excellent performance in bot attack recognition (N-day scenario), achieving an F1-score superior to 93.8\%. In the detection of unknown attacks (0-day), our solution maintains strong performance, with the exception of the eavesdropping attack. The PSO-based stacking strategy significantly enhances performance, with an average improvement of approximately 19\% in the N-day scenario and over 45\% in the 0-day scenario.


\section{Conclusion} \label{CON}
This paper proposed an edge-based Intrusion Detection System (IDS) for the Internet of Vehicles (IoV) to address the critical threats posed by bot malware and cyber attacks. By leveraging Multi-access Edge Computing (MEC) servers, our IDS monitors network traffic to and from Connected and Autonomous Vehicles (CAVs) using a meta-ensemble classifier of Isolation Forests (IFs) optimized with Particle Swarm Optimization (PSO). This approach effectively detects both known and zero-day attacks. Our extensive testing on a recent dataset of vehicular bot attacks demonstrates the IDS's high efficiency and accuracy in both recognizing N-day attacks and detecting 0-day bot attacks.



\section*{Acknowledgment}
This work was supported by the 5G-INSIGHT bilateral project (ID: 14891397) / (ANR-20-CE25-0015-16), funded by the Luxembourg National Research Fund (FNR), and by the French National Research Agency (ANR).


\bibliographystyle{unsrt}
\bibliography{ref}

\end{document}